\documentclass[preprint,showpacs,preprintnumbers,amssymb,aps,nofootinbib]{revtex4}
\usepackage{amsfonts,color,graphicx,epsfig,graphicx,amsmath,multirow,slashed}

\begin{document}

\title{The studies on $Z \to \Upsilon(1S)+g+g$ at the next-to-leading-order QCD accuracy}
\author{Zhan Sun}
\email{zhansun@cqu.edu.cn}
\affiliation{
Department of Physics, Guizhou Minzu University, Guiyang 550025, People's Republic of China.
} 

\date{\today}

\begin{abstract}
In this paper, we carry out the next-to-leading-order (NLO) studies on $Z \to \Upsilon(1S)+g+g$ via the color-singlet (CS) $b\bar{b}$ state. We find the newly calculated NLO QCD corrections to this process can significantly influence its leading-order (LO) results, and greatly improve the dependence on the renormalization scale. By including the considerable feeddown contributions, the branching ratio $\mathcal{B}_{Z \to \Upsilon(1S)+g+g}$ is predicted to be $(0.56 \sim 0.95)\times 10^{-6}$, which can reach up to $19\% \sim 31\%$ of the LO predictions given by the CS dominant process $Z \to \Upsilon(1S)+b+\bar{b}$. Moreover, $Z \to \Upsilon(1S)+g+g$ also seriously affect the CS predictions on the $\Upsilon(1S)$ energy distributions, especially when $z$ is relatively small. In summary, for the inclusive $\Upsilon(1S)$ productions in $Z$ decay, besides $Z \to \Upsilon(1S)+b+\bar{b}$, the gluon radiation process $Z \to \Upsilon(1S)+g+g$ can provide indispensable contributions as well.
\pacs{12.38.Bx, 12.39.Jh, 13.38.Dg, 14.40.Pq}

\end{abstract}

\maketitle

\section{Introduction}
The L3 group at LEP has released the measurement on the total decay width of $Z \to \Upsilon(1S)+X$ \cite{Acciarri:1998iy}
\begin{eqnarray}
\mathcal{B}_{Z \to \Upsilon(1S)+X} < 4.4 \times 10^{-5}. \nonumber
\end{eqnarray} 
The leading-order (LO) color-singlet (CS) predictions obtained by calculating the CS dominant process $Z \to \Upsilon(1S)+b+\bar{b}$ are only at the $10^{-6}$ order \cite{Li:2010xu,Barger:1989cq,Braaten:1993mp}. Subsequently Li et al. \cite{Li:2010xu} accomplish the next-to-leading-order (NLO) QCD corrections to this $b\bar{b}$ pair associated channel, pointing out the higher-order terms can give rise to a $24\%$ enhancement to the total decay width.

Reviewing the inclusive $J/\psi$ productions via $Z$ decay \cite{Alexander:1996jp,Li:2010xu,Abreu:1995ui,Guberina:1980dc,Barger:1989cq,Braaten:1993mp,Ernstrom:1996aa,Fleming:1993fq,Keung:1980ev,Boyd:1998km,Cheung:1995ka,Cho:1995vv,Baek:1996np,Lansberg:2019adr}, besides the $\alpha\alpha_s^2$-order process $Z \to J/\psi+c+\bar{c}$ that serves as the leading role in the CS predictions, the electromagnetic processes $Z \to f\bar{f}\gamma^{*}$ with $\gamma^{*} \to J/\psi$ $(f=l,u,d,s,c,b)$ and the gluon fragmentation processes $Z \to f_{q}\bar{f_{q}}g^{*}$ with $g^{*} \to J/\psi gg$ $(f_{q}=u,d,s,c,b)$ can also provide nonnegligible contributions. While, due to the suppression by $\frac{m_c^2}{m_Z^2}$ \cite{Braaten:1993mp}, the total decay width of the other $\alpha\alpha_s^2$-order process $Z \to J/\psi+g+g$ is only about two orders of magnitudes smaller than that of $Z \to J/\psi+c+\bar{c}$. As for the $\Upsilon$ productions, the situations become just the opposite. The relative significances of the electromagnetic and gluon fragmentation processes are much less important than the $J/\psi$ case, since the larger value of $m_b$ than $m_c$ highly suppress the denominator of the propagators $\gamma^{*}(\to \Upsilon)$ and $g^{*}(\to \Upsilon gg)$. However, for $Z \to \Upsilon+g+g$, the stated above suppresion effect ($\frac{m_c^2}{m_Z^2}$) will be largely weaken by replacing $m_c$ with $m_b$, subsequently making this process to be indispensable in comparison with $Z \to \Upsilon+b+\bar{b}$ \cite{Barger:1989cq}. Moreover, the $\Upsilon$ energy distributions in $Z \to \Upsilon+g+g$ and $Z \to \Upsilon+b+\bar{b}$ may be thoroughly different. This can be understood by that the former process is seriously suppressed by the factor $\frac{M_{\Upsilon}^2}{E_{\Upsilon}^2}$ for large $z$ \cite{Kuhn:1981jn,Kuhn:1981jy,Boyd:1998km}, so the value of $z$ concerning the maximum $\frac{d\Gamma}{dz}$ should be small; however, as a result of the $b$ quark fragmentation, the dominant contributions of $Z \to \Upsilon+b+\bar{b}$ exist in the region of large $z$. From these points of view, the process $Z \to \Upsilon+g+g$ would be crucial for $Z$ decaying to the inclusive $\Upsilon$, deserving a separate investigation. 

Seeing that all the existing studies on $Z \to \Upsilon(1S)+g+g$ are only accurate to the first order in $\alpha_s$, to investigate the effects of the higher-order terms, in this paper we will for the first time carry out the NLO QCD corrections to this process. In general, for the inclusive $\Upsilon(1S)$ productions, the feeddown via the excited states can provide nonnegligible contributions. Therefore, in addition to the direct productions, we will take the feeddown effects via $\Upsilon(2,3S)$ and $\chi_{bJ}(1,2,3P)$ $(J=0,1,2)$ into account as well.   

The rest of the paper is organized as follows: In Sec. II, we give a description on the calculation formalism. In Sec. III, the phenomenological results and discussions are presented. Section IV is reserved as a summary.

\section{Calculation Formalism}

\begin{figure*}
\includegraphics[width=0.95\textwidth]{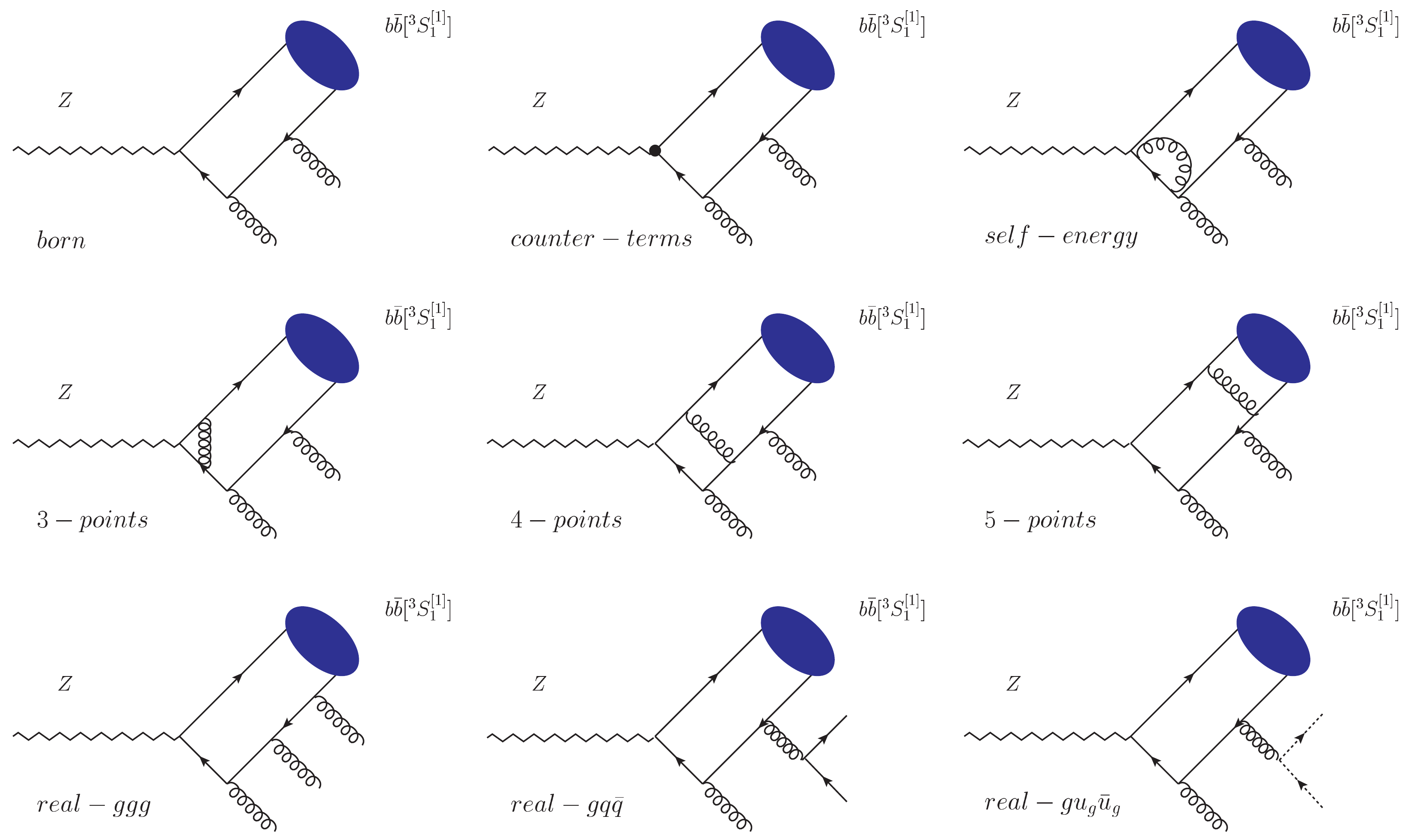}
\caption{\label{fig:Feyn}
Representative Feynman diagrams for the NLO QCD corrections to $Z \to b\bar{b}[^3S_1^{[1]}]+g+g$. The superscripts $``q"$ and $``u_{g}"$ denote the light quarks ($u,d,s$) and the ghost particles, respectively.}
\end{figure*}

Based on the nonrelativistic quantum chromodynamics \cite{Bodwin:1994jh}, the decay width of $Z \to \Upsilon(\chi_{bJ})+g+g$\footnote{For the final gluons associated $\Upsilon$($\chi_{bJ}$) productions in $Z$ decay via the CS $b\bar{b}$ states, due to the color conservation, the process $Z \to b\bar{b}[^3S_1^{[1]}]+g$ is forbidden, so the lowest order process in $\alpha_s$ is $Z \to b\bar{b}[^3S_1^{[1]},^3P_J^{[1]}]+g+g$.} can be factorized as
\begin{eqnarray}
\Gamma=\hat{\Gamma}_{Z \to b\bar{b}[n]+g+g}\langle \mathcal O ^{\Upsilon(\chi_{bJ})}(n)\rangle,
\end{eqnarray}
where $\hat{\Gamma}_{Z \to b\bar{b}[n]+g+g}$ is the perturbative calculable short distance coefficients, representing the production of a configuration of the $b\bar{b}[n]$ intermediate state. The universal nonperturbative long distance matrix elements (LDMEs) $\langle \mathcal O ^{\Upsilon(\chi_{bJ})}(n)\rangle$ stands for the probability of $b\bar{b}[n]$ into $\Upsilon(\chi_{bJ})$. In our studies, we only concentrate on the CS contributions, so $n=^3S_1^{[1]}$ for $\Upsilon$, and $n=^3P_J^{[1]}$ ($J=0,1,2$) for $\chi_{bJ}$. The procedures for dealing with the soft singularities involved in $Z \to b\bar{b}[^3P_J^{[1]}]+g+g$ ($J=0,1,2$) have been described detailedly in our previours paper \cite{Sun:2019cxx,Sun:2018hpb}, so here we just give a brief presentation on the $b\bar{b}[^3S_1^{[1]}]$ related calculations.

The NLO short distance coefficients for $n=^3S_1^{[1]}$ can be written as
\begin{eqnarray}
\hat{\Gamma}^{NLO}_{Z \to b\bar{b}[^3S_1^{[1]}]+g+g}=\hat{\Gamma}_{\textrm{Born}}+\hat{\Gamma}_{\textrm{Virtual}}+\hat{\Gamma}_{\textrm{Real}},
\end{eqnarray}
where
\begin{eqnarray}
&&\hat{\Gamma}_{\textrm{Virtual}}=\hat{\Gamma}_{\textrm{Loop}}+\hat{\Gamma}_{\textrm{CT}}, \nonumber \\
&&\hat{\Gamma}_{\textrm{Real}}=\hat{\Gamma}_{\textrm{S}}+\hat{\Gamma}_{\textrm{HC}}+\hat{\Gamma}_{\textrm{H}\overline{\textrm{C}}}. \label{channel}
\end{eqnarray}
$\hat{\Gamma}_{\textrm{Virtual}}$ is the virtual corrections, consisting of the contributions from the one-loop diagrams ($\hat{\Gamma}_{\textrm{Loop}}$) and the counter terms ($\hat{\Gamma}_{\textrm{CT}}$). $\hat{\Gamma}_{\textrm{Real}}$ stands for the real corrections, which include the soft terms ($\hat{\Gamma}_{S}$), hard-collinear terms $(\hat{\Gamma}_{\textrm{HC}})$, and hard-noncollinear terms $(\hat{\Gamma}_{\textrm{H}\overline{\textrm{C}}})$. For $\hat{\Gamma}_{\textrm{Real}}$, three processes are involved:
\begin{eqnarray}
Z & \to & b\bar{b}[^3S_1^{[1]}]+g+g+g, \nonumber \\
Z & \to & b\bar{b}[^3S_1^{[1]}]+g+q+\bar{q}~(q=u,d,s), \nonumber \\
Z & \to & b\bar{b}[^3S_1^{[1]}]+g+u_g+\bar{u}_g~(\textrm{ghost}).
\end{eqnarray}
There are 177 Feynman diagrams in total, including 6 diagrams for $\hat{\Gamma}_{\textrm{Born}}$, 111 diagrams for $\hat{\Gamma}_{\textrm{Virtual}}$ (30 counter-terms, 12 self-energy, 30 3-points, 27 4-points, 12 5-points), and 60 diagrams for $\hat{\Gamma}_{\textrm{Real}}$ (42 $ggg_{\rm{v}}$, 6 $ggg_{\rm{av}}$, 6 $gq\bar{q}$, and 6 $gu_g\bar{u}_g$), as representatively shown in Fig. \ref{fig:Feyn}. $ggg_{\rm{v}}$ and $ggg_{\rm{av}}$ denote the vector and axial-vector parts of $Z \to b\bar{b}[^3S_1^{[1]}]+g+g+g$, respectively.

To isolate the ultraviolet (UV) and infrared (IR) divergences, we adopt the dimensional regularization with $D=4-2\epsilon$. The on-mass-shell (OS) scheme is employed to set the renormalization constants for the heavy quark mass ($Z_m$), heavy quark filed ($Z_2$), and gluon filed ($Z_3$). The modified minimal-subtraction ($\overline{MS}$) scheme is used for the QCD gauge coupling ($Z_g$). The renormalization constants are \cite{Sun:2018hpb,Sun:2019cxx,Gong:2012ah}
\begin{eqnarray}
\delta Z_{m}^{OS}&=& -3 C_{F} \frac{\alpha_s N_{\epsilon}}{4\pi}\left[\frac{1}{\epsilon_{\textrm{UV}}}-\gamma_{E}+\textrm{ln}\frac{4 \pi \mu_r^2}{m_b^2}+\frac{4}{3}\right], \nonumber \\
\delta Z_{2}^{OS}&=& - C_{F} \frac{\alpha_s N_{\epsilon}}{4\pi}\left[\frac{1}{\epsilon_{\textrm{UV}}}+\frac{2}{\epsilon_{\textrm{IR}}}-3 \gamma_{E}+3 \textrm{ln}\frac{4 \pi \mu_r^2}{m_b^2}+4\right], \nonumber \\
\delta Z_{3}^{OS}&=& \frac{\alpha_s N_{\epsilon}}{4\pi}\left[(\beta_{0}^{'}-2 C_{A})(\frac{1}{\epsilon_{\textrm{UV}}}-\frac{1}{\epsilon_{\textrm{IR}}})-\frac{4}{3}T_F(\frac{1}{\epsilon_{\textrm{UV}}}-\gamma_E+\textrm{ln}\frac{4\pi\mu_r^2}{m_c^2}) \right. \nonumber\\
&& \left. -\frac{4}{3}T_F(\frac{1}{\epsilon_{\textrm{UV}}}-\gamma_E+\textrm{ln}\frac{4\pi\mu_r^2}{m_b^2})\right], \nonumber \\
\delta Z_{g}^{\overline{MS}}&=& -\frac{\beta_{0}}{2}\frac{\alpha_s N_{\epsilon}}{4\pi}\left[\frac{1} {\epsilon_{\textrm{UV}}}-\gamma_{E}+\textrm{ln}(4\pi)\right], \label{CT}
\end{eqnarray}
where $\gamma_E$ is the Euler's constant, $N_{\epsilon}= \Gamma[1-\epsilon] /({4\pi\mu_r^2}/{(4m_b^2)})^{\epsilon}$, $\beta_{0}=\frac{11}{3}C_A-\frac{4}{3}T_Fn_f$ is the one-loop coefficient of the $\beta$-function, and $\beta_{0}^{'}=\frac{11}{3}C_A-\frac{4}{3}T_Fn_{lf}$. $n_f(=5)$ and $n_{lf}(=n_f-2)$ are the number of active quark flavors and light quark flavors, respectively. In ${\rm SU}(3)$, the color factors are given by $T_F=\frac{1}{2}$, $C_F=\frac{4}{3}$, and $C_A=3$. The two-cutoff slicing strategy is utilized to subtract the IR divergences in $\Gamma_{\textrm{Real}}$ \cite{Harris:2001sx}.

\begin{figure*}
\includegraphics[width=0.495\textwidth]{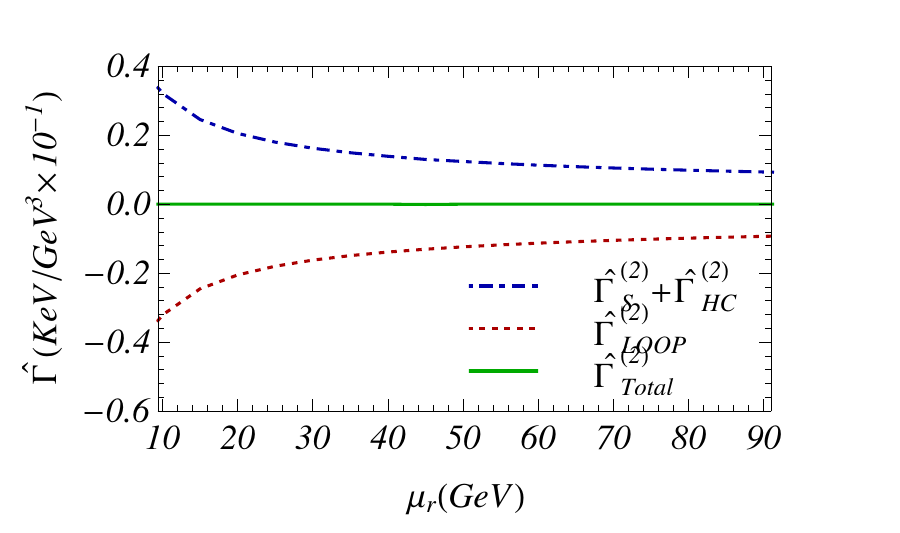}
\includegraphics[width=0.495\textwidth]{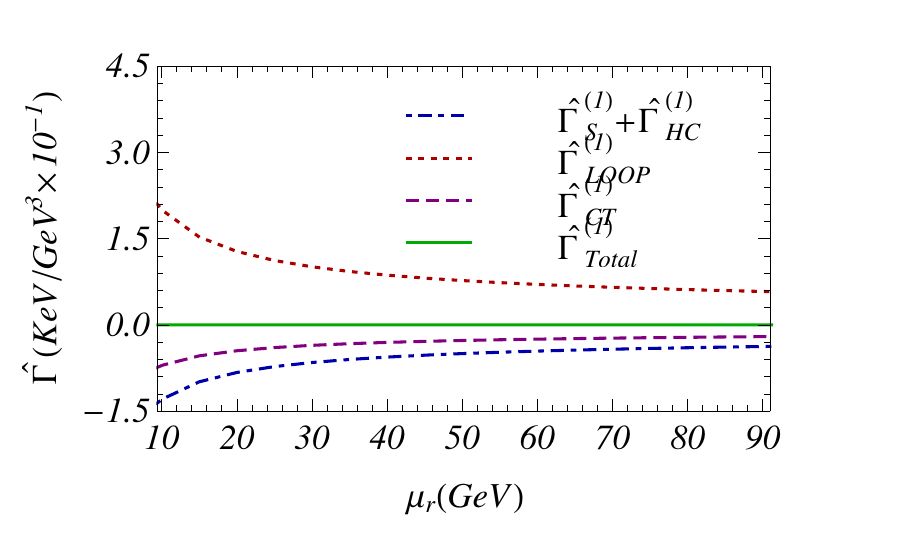}
\caption{\label{fig:divergence}
Cancellation of the $\epsilon^{-2}-$ and $\epsilon^{-1}-$order divergences. The superscripts $``(2)"$ and $``(1)"$ denote the $\epsilon^{-2}-$ and $\epsilon^{-1}-$order terms, respectively.}
\end{figure*}

\begin{figure*}
\includegraphics[width=0.495\textwidth]{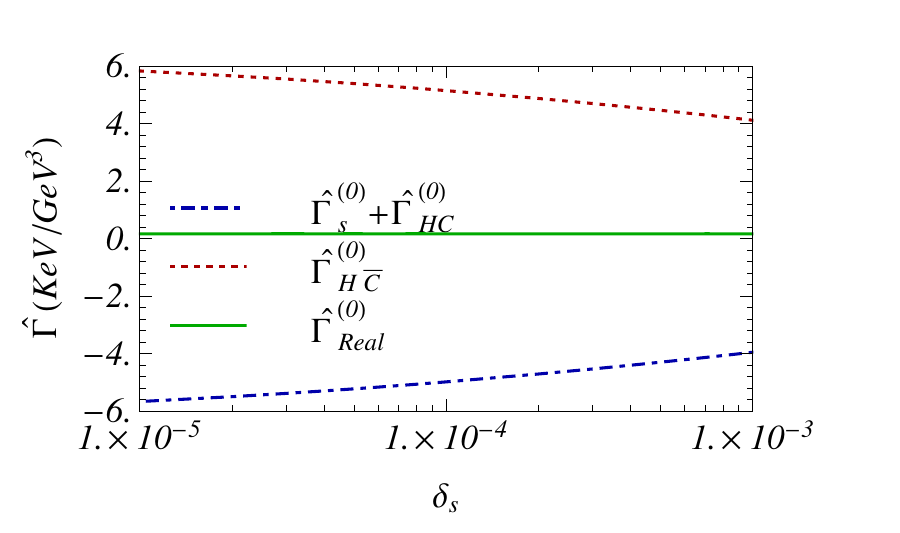}
\includegraphics[width=0.495\textwidth]{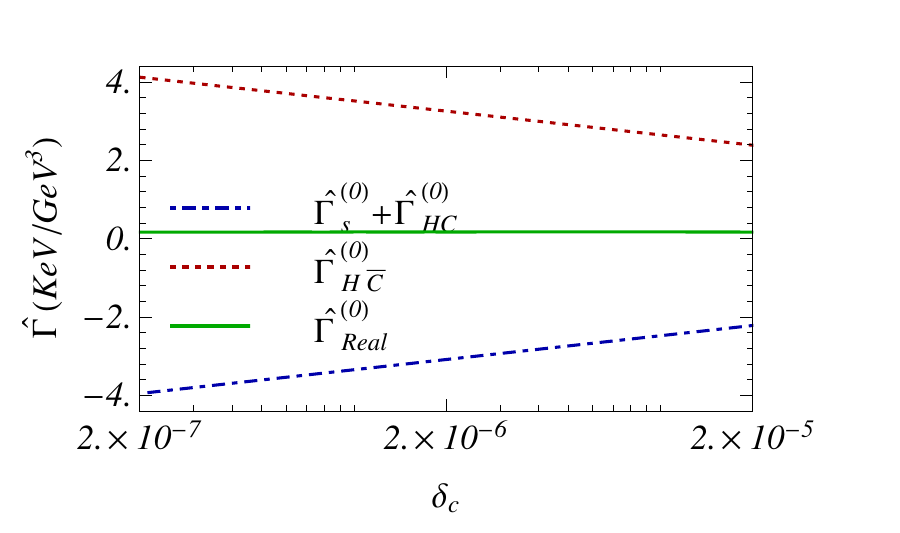}
\caption{\label{fig:cut}
The verification of the independence on the cutoff parameters $\delta_s$ and $\delta_c$. The superscript $``(0)"$ denotes the $\epsilon^{0}-$order terms. In the left diagram, $\delta_c=2 \times 10^{-7}$. In the right diagram, $\delta_s=1 \times 10^{-3}$.}
\end{figure*}

The package $\textrm{M}\scriptsize{\textrm{ALT@FDC}}$ that has been employed to preform the NLO QCD corrections to several heavy quarkonium related processes \cite{Sun:2017wxk,Sun:2018rgx,Sun:2018hpb,Sun:2019cxx,Jiang:2018wmv} is used to deal with $\hat{\Gamma}_{\textrm{Virtual}}$, $\hat{\Gamma}_{\textrm{\textrm{S}}}$, and $\hat{\Gamma}_{\textrm{\textrm{HC}}}$. The $\textrm{FDC}$ package \cite{Wang:2004du} serves as the agent for calculating the hard-noncollinear part $\hat{\Gamma}_{\textrm{H}\overline{\textrm{C}}}$. Both the cancellation of the $\epsilon^{-2(-1)}$-order divergences and the independence on the cutoff parameters $\delta_{s,c}$ have been checked carefully, as shown in Figs. \ref{fig:divergence} and \ref{fig:cut}.

As a crosscheck, taking the same input parameters, we have reproduced the  NLO results of $\sigma_{e^+e^- \to J/\psi+g+g}$ in Refs. \cite{Ma:2008gq,Gong:2009kp}.

\section{Phenomenological results}
In our calculations, the input parameters are taken as
\begin{eqnarray}
&&\alpha=1/128,~~~m_b=4.9~\textrm{GeV},~~~m_c=1.5~\textrm{GeV},\nonumber \\
&&m_Z=91.1876~\textrm{GeV},~~~m_{q/\bar{q}}=0~(q=u,d,s),\nonumber \\
&&\sin^{2}(\theta_W)=0.23116. \label{para}
\end{eqnarray}
To determine $\langle \mathcal O ^{\Upsilon(nS)}(^3S_1^{[1]})\rangle$ and $\langle \mathcal O ^{\chi_{bJ}(mP)}(^3P_J^{[1]})\rangle$, we employ the relations to the radial wave functions at the origin ($n,m=1,2,3$)
\begin{eqnarray}
\frac{\langle \mathcal O^{\Upsilon(nS)}(^3S_1^{[1]}) \rangle}{6N_c}&=&\frac{1}{4\pi}|R_{\Upsilon(nS)}(0)|^2, \\ \nonumber
\frac{\langle \mathcal O^{\chi_{bJ}(mP)}(^3P_J^{[1]}) \rangle}{2N_c}&=&(2J+1)\frac{3}{4\pi}|R^{'}_{\chi_b(mP)}(0)|^2,
\end{eqnarray}
where $|R_{\Upsilon(nS)}(0)|^2$ and $|R^{'}_{\chi_b(mP)}(0)|^2$ read \cite{Eichten:1995ch}
\begin{eqnarray}
&&|R_{\Upsilon(1S)}(0)|^2=6.477~\textrm{GeV}^3,~~~|R_{\Upsilon(2S)}(0)|^2=3.234~\textrm{GeV}^3, \\ \nonumber
&&|R_{\Upsilon(3S)}(0)|^2=2.474~\textrm{GeV}^3, \\ \nonumber
&&|R^{'}_{\chi_b(1P)}(0)|^2=1.417~\textrm{GeV}^5,~~~|R^{'}_{\chi_b(2P)}(0)|^2=1.653~\textrm{GeV}^5, \\ \nonumber
&&|R^{'}_{\chi_b(3P)}(0)|^2=1.794~\textrm{GeV}^5. \nonumber
\end{eqnarray}
Branching ratios of $\chi_{bJ}(mP) \to \Upsilon(nS)$, $\Upsilon(nS) \to \chi_{bJ}(mP)$, $\Upsilon(3S) \to \Upsilon(2S)$, $\Upsilon(3S) \to \Upsilon(1S)$, and $\Upsilon(2S) \to \Upsilon(1S)$ can be found in Refs. \cite{Gong:2013qka,Han:2014kxa,Feng:2015wka}.

\begin{table*}[htb]
\caption{The total decay widths of $Z \to \Upsilon(1S)+g+g$ (in units of $\textrm{KeV}$). $K$ denotes $\Gamma^{NLO}_{DR}/\Gamma^{LO}_{DR}$. $``NLO"$ represents the sum of the contribution of the LO terms and that of the QCD corrections. The superscripts $``DR"$ and $``FD"$ denote the direct and feeddown contributions, respectively.}
\label{decay width}
\begin{tabular}{ccccccccc}
\hline\hline
$\mu_r$ & $m_b$ (GeV) & $~\Gamma^{LO}_{DR}~$ & $~\Gamma^{NLO}_{DR}~$ & $~K~$ & $~\Gamma_{FD}^{\chi_{bJ}(1,2,3P)}~$ & $~\Gamma_{FD}^{\Upsilon(2,3S)}~$ & $~\Gamma_{Total}~$\\ \hline
$~$ & $4.7$ & $1.94$ & $1.83$ & $0.94$ & $0.20$ & $0.34$ & $2.37$\\
$2m_b$ & $4.9$ & $1.75$ & $1.66$ & $0.95$ & $0.17$ & $0.30$ & $2.13$\\
$~$ & $5.1$ & $1.59$ & $1.51$ & $0.95$ & $0.15$ & $0.28$ & $1.94$\\ \hline
$~$ & $4.7$ & $0.82$ & $1.33$ & $1.62$ & $0.10$ & $0.24$ & $1.67$\\
$m_Z$ & $4.9$ & $0.76$ & $1.22$ & $1.61$ & $0.09$ & $0.22$ & $1.53$\\
$~$ & $5.1$ & $0.70$ & $1.12$ & $1.60$ & $0.08$ & $0.20$ & $1.40$\\ \hline\hline
\end{tabular}
\end{table*}

\begin{figure*}
\includegraphics[width=0.65\textwidth]{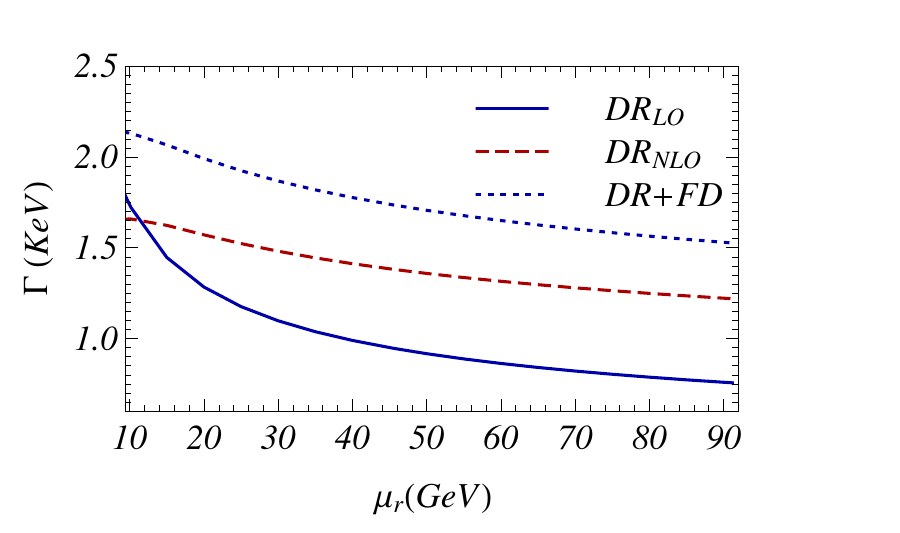}
\caption{\label{fig:mur}
The total decay widths of $Z \to \Upsilon(1S)+g+g$ as a function of the renormalization scale $\mu_r$ with $m_b=4.9$ GeV. $``NLO"$ represents the sum of the contribution of the LO terms and that of the QCD corrections. The superscripts $``DR"$ and $``FD"$ denote the direct and feeddown contributions, respectively.}
\end{figure*}

The total decay widths of $Z \to \Upsilon(1S)+g+g$ are listed in Table. \ref{decay width}. To demonstrate the dependence on the renormalization scale $\mu_r$, the results for $\mu_r=2m_b$ and $\mu_r=m_Z$ are presented simultaneously. From the data in this table, one can observe
\begin{itemize}
\item[i)]
For the direct productions, when $\mu_r=2m_b$, the QCD corrections diminish the LO results by about $5\%$, and cause a $60\%$ enhancement for $\mu_r=m_Z$. In addition, incorporating these higher-order terms notably weaken the dependence on $\mu_r$. As is illustrated in Fig. \ref{fig:mur}, the line referring to ``$DR_{NLO}$" decreases much more slowly than that for ``$DR_{LO}$" with the increase of $\mu_r$.
\item[ii)]
The decay of $\Upsilon(2,3S)$ and $\chi_{bJ}(1,2,3P)$ can raise the predictions of the direct productions by about $30\%$, manifestly indicating the feeddown significance in the $\Upsilon(1S)$ production.
\item[iii)]
The dependence on the mass of the $b$ quark is mild, e.g., varying $m_b$ by $\pm~0.2$ GeV from the central value of 4.9 GeV just results in a $10\%$ variation of the total decay width.
\end{itemize}

Summing up the direct and feeddown contributions, we finally obtain
\begin{eqnarray}
\mathcal{B}_{Z \to \Upsilon(1S)+g+g} = (0.56 \sim 0.95) \times 10^{-6},
\end{eqnarray}
where the theoretical uncertainty is induced by the choices of the values of $\mu_r$ ($2m_b \sim m_Z$) and $m_b$ ($4.9 \sim 5.1$ GeV).

\begin{figure*}
\includegraphics[width=0.495\textwidth]{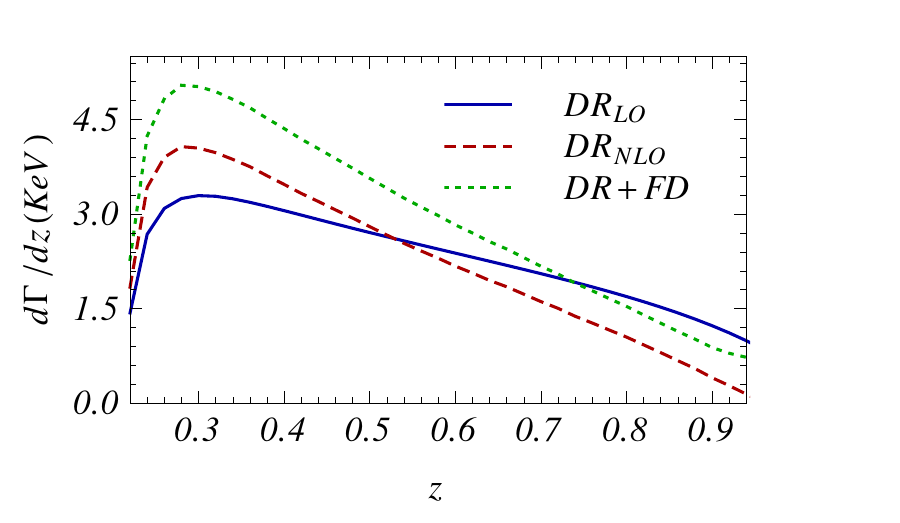}
\includegraphics[width=0.495\textwidth]{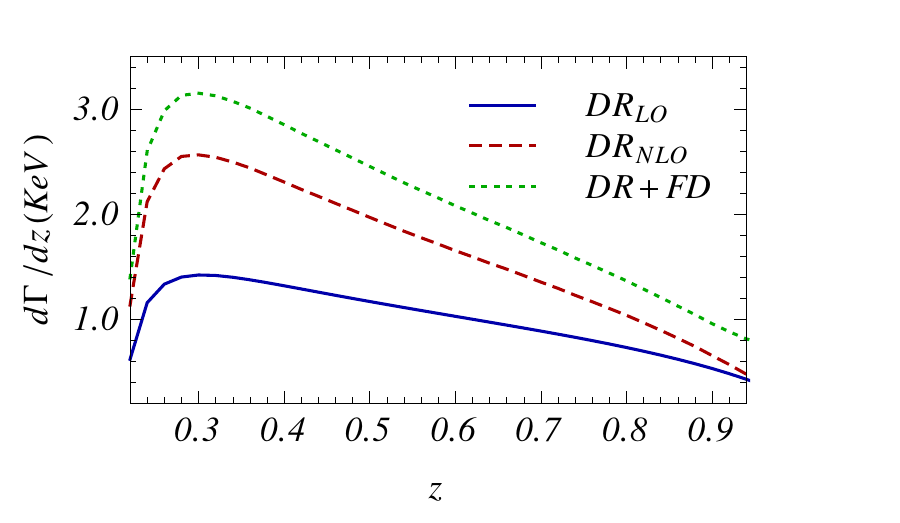}
\caption{\label{fig:zdis}
The $\Upsilon(1S)$ energy distributions in $Z \to \Upsilon(1S)+g+g$. $z=\frac{2E_{\Upsilon(1S)}}{m_Z}$ and $m_b=4.9$ GeV. $``NLO"$ represents the sum of the contribution of the LO terms and that of the QCD corrections. $\mu_r=2m_b$ in the left diagram, and $\mu_r=m_Z$ in the right diagram. The superscripts $``DR"$ and $``FD"$ denote the direct and feeddown contributions, respectively.}
\end{figure*}

In Fig. \ref{fig:zdis}, the $\Upsilon(1S)$ energy distributions in $Z \to \Upsilon(1S)+g+g$ are presented with $z$ defined as $\frac{2E_{\Upsilon(1S)}}{m_Z}$. For the direct productions, when $\mu_r$ is equal to $2m_b$, the QCD corrections are positive for $z<0.52$, and negative in the remaining scope of $z$; however, for $\mu_r=m_Z$, these corrections keep always positive for all the available $z$ values. With $\mu_r$ being relatively small, the higher-order terms can significantly enhance the differential decay width. Taking $z=0.3$ as an example,
\begin{eqnarray}
\left(\frac{d\Gamma}{dz}\right)_{NLO} / \left(\frac{d\Gamma}{dz}\right)_{LO}&=&1.23~\textrm{for}~\mu_r=2m_b, \nonumber \\
\left(\frac{d\Gamma}{dz}\right)_{NLO} / \left(\frac{d\Gamma}{dz}\right)_{LO}&=&1.81~\textrm{for}~\mu_r=m_Z,
\end{eqnarray}
where $``NLO"$ represents the sum of the contribution of the LO terms and that of the QCD corrections. Including the feedown contributions can further increase $\frac{d\Gamma}{dz}$ by about $20\% \sim 30\%$ for most values of $z$. Note that, phenomenological and theoretical arguments suggest that the NRQCD factorization holds only when the quarkonium is produced at relatively large momentum, thus in Fig. \ref{fig:zdis} the predicted $z$ distributions for very small $z$ values (close to the left endpoint) may not hold.  

\begin{figure*}
\includegraphics[width=0.495\textwidth]{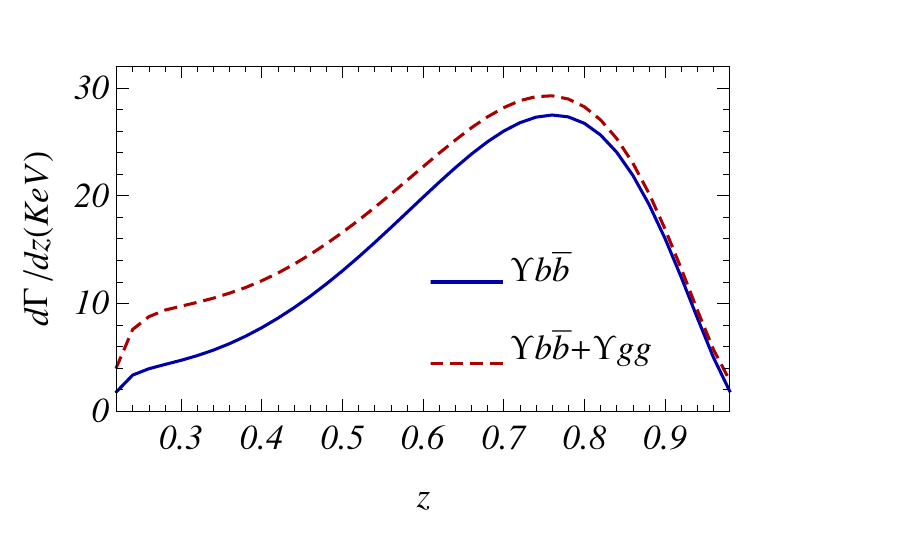}
\includegraphics[width=0.495\textwidth]{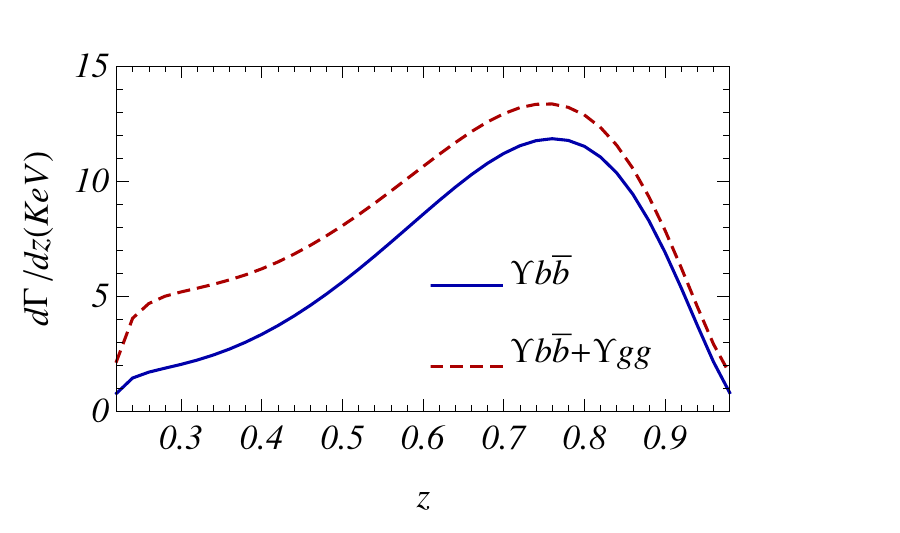}
\caption{\label{fig:zdiscompare}
The differential decay widths. $z=\frac{2E_{\Upsilon(1S)}}{m_Z}$ and $m_b=4.9$ GeV. $\mu_r=2m_b$ in the left diagram, and $\mu_r=m_Z$ in the right diagram. The superscripts $``\Upsilon gg"$ and $``\Upsilon b\bar{b}"$ denote the contributions via $Z \to \Upsilon(1S)+g+g$ and $Z \to \Upsilon(1S)+b+\bar{b}$, respectively.}
\end{figure*}

At last, we simply compare the contributions of $Z \to \Upsilon(1S)+g+g$ with that of the dominant channel $Z \to \Upsilon(1S)+b+\bar{b}$. With the help of the $FDC$ package, taking the parameters in Eq. \ref{para}, we have 
\begin{eqnarray}
\Gamma^{LO,\mu_r=2m_b}_{Z \to \Upsilon(1S)+b+\bar{b}} & = & 11.5~KeV, \nonumber \\
\Gamma^{LO,\mu_r=m_Z}_{Z \to \Upsilon(1S)+b+\bar{b}} & = & 4.98~KeV.
\end{eqnarray}
Comparing to the results with $m_b=4.9$ GeV in Tab. \ref{decay width}, one can obtain 
\begin{eqnarray}
\frac{\Gamma_{Z \to \Upsilon(1S)+g+g}}{\Gamma^{LO}_{Z \to \Upsilon(1S)+b+\bar{b}}} = 19\% \sim 31\%,
\end{eqnarray}
where the uncertainty arises from the variation of $\mu_r$ in $\left[2m_b,m_Z\right]$. This ratio suggests that, for the total decay width, the contributions via $Z \to \Upsilon(1S)+g+g$ is comparable with the magnitude of the QCD corrections to $Z \to \Upsilon(1S)+b+\bar{b}$. In Fig. \ref{fig:zdiscompare}, the comparison of the $\Upsilon(1S)$ energy distributions in $Z \to \Upsilon(1S)+g+g$ and $Z \to \Upsilon(1S)+b+\bar{b}$ are presented, where $``\Upsilon gg"$ denotes the contributions via the former process up to the NLO accuracy in $\alpha_s$, including the feeddown effects; $``\Upsilon b\bar{b}"$ stands for the direct contributions of the $b\bar{b}$ pair associated process at the LO QCD accuracy. One can find, for relatively small $z$, adding the $\Upsilon gg$ contributions can increase the $\Upsilon b\bar{b}$ predictions to a surprisingly large extent. Such a remarkable enhancement on $\frac{d\Gamma}{dz}$ is almost the same in size as the QCD corrections to $Z \to \Upsilon(1S)+b+\bar{b}$. All these points obviously reveal the phenomenological  importance of $Z \to \Upsilon(1S)+g+g$ for $Z$ decaying to the inclusive $\Upsilon(1S)$. 

Note that, within the NRQCD framework, for the $\Upsilon$ production in association with final gluon(s) via $Z$ decay, in addition to the CS process $Z \to b\bar{b}[^3S_1^{[1]}]+g+g$ that we focus on in our present paper, the color-octet (CO) channels $Z \to b\bar{b}[^1S_0^{[8]},^3S_1^{[8]},^3P_J^{[8]}]+g$ are also allowed. Generally speaking, according to the velocity-scaling rule of NRQCD, the CO LDMEs are suppressed by a power of $v^{4}$ ($v^2_{b} \simeq 0.1$), where $v$ denotes the relative velocity between the constituent quark and anti-quark of the heavy quarkonium. However, the less of an $\alpha_s$ in $Z \to b\bar{b}[^1S_0^{[8]},^3S_1^{[8]},^3P_J^{[8]}]+g$ comparing to $Z \to b\bar{b}[^3S_1^{[1]}]+g+g$, together with the potential enhancement via the NLO corrections to these CO channels\footnote{Such as the fragmentation process of $Z \to q\bar{q}g^{*}$ with $g^{*}\to b\bar{b}[^3S_1^{[8]}]$.}, may partly compensate for the suppression caused by LDMEs, subsequently making the CO contributions nonnegligible. Of course, whether this is indeed the case depends on the future rigorous NLO calculations for these CO production channels.

\section{Summary}
In this manuscript, we for the first time perform the complete NLO studies on the process $Z \to \Upsilon(1S)+g+g$ via the CS $b\bar{b}$ states. We find the impacts of the QCD corrections on the LO results are significant, including both the total and differential decay widths. In addition, these higher-order terms markedly weaken the dependence of the theoretical predictions on $\mu_r$. By incorporating the substantial feeddown effects via $\Upsilon(2,3S)$ and $\chi_{bJ}(1,2,3P)$, $\mathcal{B}_{Z \to \Upsilon(1S)+g+g}$ is scattered in the range $(0.56 \sim 0.95) \times 10^{-6}$. This value is about $19\% \sim 31\%$ of the LO predictions given by $Z \to \Upsilon(1S)+b+\bar{b}$, which is responsible for the main contributions in the CS predictions. Moreover, for small $E_{\Upsilon(1S)}$, $Z \to \Upsilon(1S)+g+g$ has vital influence on the $\Upsilon(1S)$ energy distributions. In view of these points, for $Z$ decaying to the inclusive $\Upsilon(1S)$, in addition to $Z \to \Upsilon(1S)+b+\bar{b}$, the process $Z \to \Upsilon(1S)+g+g$ is also phenomenologically important. 
\section{Acknowledgments}
\noindent{\bf Acknowledgments}:
This work is supported in part by the Natural Science Foundation of China under the Grant No. 11647113. and No. 11705034., by the Project for Young Talents Growth of Guizhou Provincial Department of Education under Grant No.KY[2017]135, and by the Project of GuiZhou Provincial Department of Science and Technology under Grant No. QKHJC[2019]1160.\\


\begin{thebibliography}{1}

\bibitem{Acciarri:1998iy}
  M.~Acciarri {\it et al.} [L3 Collaboration],
  ``Heavy quarkonium production in $Z$ decays,''
  Phys.\ Lett.\ B {\bf 453} (1999) 94.
  doi:10.1016/S0370-2693(99)00280-4.

\bibitem{Barger:1989cq}
  V.~D.~Barger, K.~m.~Cheung and W.~Y.~Keung,
  ``Z Boson Decays To Heavy Quarkonium,''
  Phys.\ Rev.\ D {\bf 41} (1990) 1541.
  doi:10.1103/PhysRevD.41.1541.

\bibitem{Braaten:1993mp}
  E.~Braaten, K.~m.~Cheung and T.~C.~Yuan,
  ``Z0 decay into charmonium via charm quark fragmentation,''
  Phys.\ Rev.\ D {\bf 48} (1993) 4230
  doi:10.1103/PhysRevD.48.4230.

\bibitem{Li:2010xu}
  R.~Li and J.~X.~Wang,
  ``The next-to-leading-order QCD correction to inclusive $J/\psi(\Upsilon)$ production in $Z^0$ decay,''
  Phys.\ Rev.\ D {\bf 82} (2010) 054006
  doi:10.1103/PhysRevD.82.054006.

\bibitem{Alexander:1996jp}
  G.~Alexander {\it et al.} [OPAL Collaboration],
  ``Prompt J / psi production in hadronic Z0 decays,''
  Phys.\ Lett.\ B {\bf 384} (1996) 343.
  doi:10.1016/0370-2693(96)00656-9.

\bibitem{Abreu:1995ui}
  P.~Abreu {\it et al.} [DELPHI Collaboration],
  ``Search for promptly produced heavy quarkonium states in hadronic Z decays,''
  Z.\ Phys.\ C {\bf 69} (1996) 575.
  doi:10.1007/s002880050061.

\bibitem{Guberina:1980dc}
  B.~Guberina, J.~H.~Kuhn, R.~D.~Peccei and R.~Ruckl,
  ``Rare Decays of the Z0,''
  Nucl.\ Phys.\ B {\bf 174} (1980) 317.
  doi:10.1016/0550-3213(80)90287-4.

\bibitem{Ernstrom:1996aa}
  P.~Ernstrom, L.~Lonnblad and M.~Vanttinen,
  ``Evolution effects in $Z^0$ fragmentation into charmonium,''
  Z.\ Phys.\ C {\bf 76} (1997) 515
  doi:10.1007/s002880050574.

\bibitem{Fleming:1993fq}
  S.~Fleming,
  ``Electromagnetic production of quarkonium in Z0 decay,''
  Phys.\ Rev.\ D {\bf 48} (1993) R1914
  doi:10.1103/PhysRevD.48.R1914.

\bibitem{Keung:1980ev}
  W.~Y.~Keung,
  ``Off Resonance Production of Heavy Vector Quarkonium States in $e^+ e^-$ Annihilation,''
  Phys.\ Rev.\ D {\bf 23} (1981) 2072.
  doi:10.1103/PhysRevD.23.2072.

\bibitem{Boyd:1998km}
  C.~G.~Boyd, A.~K.~Leibovich and I.~Z.~Rothstein,
  ``J / psi production at LEP: Revisited and resummed,''
  Phys.\ Rev.\ D {\bf 59} (1999) 054016
  doi:10.1103/PhysRevD.59.054016.

\bibitem{Cheung:1995ka}
  K.~m.~Cheung, W.~Y.~Keung and T.~C.~Yuan,
  ``Color octet quarkonium production at the $Z$ pole,''
  Phys.\ Rev.\ Lett.\  {\bf 76} (1996) 877
  doi:10.1103/PhysRevLett.76.877.

\bibitem{Cho:1995vv}
  P.~L.~Cho,
  ``Prompt upsilon and psi production at LEP,''
  Phys.\ Lett.\ B {\bf 368} (1996) 171
  doi:10.1016/0370-2693(95)01484-5.

\bibitem{Baek:1996np}
  S.~Baek, P.~Ko, J.~Lee and H.~S.~Song,
  ``Color octet heavy quarkonium productions in Z0 decays at LEP,''
  Phys.\ Lett.\ B {\bf 389} (1996) 609
  doi:10.1016/S0370-2693(96)01313-5.

\bibitem{Lansberg:2019adr}
  J.~P.~Lansberg,
  ``New Observables in Inclusive Production of Quarkonia,''
  arXiv:1903.09185 [hep-ph].
  
\bibitem{Kuhn:1981jn}
  J.~H.~Kuhn and H.~Schneider,
  ``Inclusive $J/\psi^\prime$s in $e^+ e^-$ Annihilations,''
  Phys.\ Rev.\ D {\bf 24} (1981) 2996.
  doi:10.1103/PhysRevD.24.2996.

\bibitem{Kuhn:1981jy}
  J.~H.~Kuhn and H.~Schneider,
  ``Testing {QCD} Through Inclusive $J/\psi$ Production in $e^+ e^-$ Annihilations,''
  Z.\ Phys.\ C {\bf 11} (1981) 263.
  doi:10.1007/BF01545683.

\bibitem{Bodwin:1994jh}
  G.~T.~Bodwin, E.~Braaten and G.~P.~Lepage,
  ``Rigorous QCD analysis of inclusive annihilation and production of heavy quarkonium,''
  Phys.\ Rev.\ D {\bf 51} (1995) 1125
   Erratum: [Phys.\ Rev.\ D {\bf 55} (1997) 5853]
  doi:10.1103/PhysRevD.55.5853, 10.1103/PhysRevD.51.1125.

\bibitem{Sun:2019cxx}
  Z.~Sun and Y.~Ma,
  ``Inclusive productions of $\Upsilon(1S,2S,3S)$ and $\chi_b(1P,2P,3P)$ via the Higgs boson decay,''
  Phys.\ Rev.\ D {\bf 100} (2019) no.9,  094019
  doi:10.1103/PhysRevD.100.094019.

\bibitem{Sun:2018hpb}
  Z.~Sun and H.~F.~Zhang,
  ``Next-to-leading-order QCD corrections to the decay of $Z$ boson into $\chi_c(\chi_b)$,''
  Phys.\ Rev.\ D {\bf 99} (2019) no.9,  094009
  doi:10.1103/PhysRevD.99.094009.

\bibitem{Gong:2012ah}
  B.~Gong, J.~P.~Lansberg, C.~Lorce and J.~Wang,
  ``Next-to-leading-order QCD corrections to the yields and polarisations of J/Psi and Upsilon directly produced in association with a Z boson at the LHC,''
  JHEP {\bf 1303} (2013) 115
  doi:10.1007/JHEP03(2013)115.

\bibitem{Harris:2001sx}
  B.~W.~Harris and J.~F.~Owens,
  ``The Two cutoff phase space slicing method,''
  Phys.\ Rev.\ D {\bf 65} (2002) 094032
  doi:10.1103/PhysRevD.65.094032. 

\bibitem{Sun:2017wxk}
  Z.~Sun and H.~F.~Zhang,
  ``QCD corrections to the color-singlet $J/\psi$ production in deeply inelastic scattering at HERA,''
  Phys.\ Rev.\ D {\bf 96} (2017) no.9,  091502
  doi:10.1103/PhysRevD.96.091502.
  
\bibitem{Sun:2018rgx}
  Z.~Sun, X.~G.~Wu, Y.~Ma and S.~J.~Brodsky,
  ``Exclusive production of $J/\psi+\eta_c$ at the $B$ factories Belle and Babar using the principle of maximum conformality,''
  Phys.\ Rev.\ D {\bf 98} (2018) no.9,  094001
  doi:10.1103/PhysRevD.98.094001.

\bibitem{Jiang:2018wmv}
  Y.~Jiang and Z.~Sun,
  Eur.\ Phys.\ J.\ C {\bf 78} (2018) no.11,  892
  doi:10.1140/epjc/s10052-018-6392-x.
  
\bibitem{Wang:2004du}
  J.~X.~Wang,
  ``Progress in FDC project,''
  Nucl.\ Instrum.\ Meth.\ A {\bf 534} (2004) 241
  doi:10.1016/j.nima.2004.07.094.

\bibitem{Ma:2008gq}
  Y.~Q.~Ma, Y.~J.~Zhang and K.~T.~Chao,
  ``QCD correction to $e^+e^- \to J/\psi+gg$ at B Factories,''
  Phys.\ Rev.\ Lett.\  {\bf 102} (2009) 162002
  doi:10.1103/PhysRevLett.102.162002.

\bibitem{Gong:2009kp}
  B.~Gong and J.~X.~Wang,
  ``Next-to-Leading-Order QCD Corrections to e+ e- ---> J/psi gg at the B Factories,''
  Phys.\ Rev.\ Lett.\  {\bf 102} (2009) 162003
  doi:10.1103/PhysRevLett.102.162003. 
  
\bibitem{Eichten:1995ch}
  E.~J.~Eichten and C.~Quigg,
  ``Quarkonium wave functions at the origin,''
  Phys.\ Rev.\ D {\bf 52} (1995) 1726
  doi:10.1103/PhysRevD.52.1726.
  
\bibitem{Gong:2013qka}
  B.~Gong, L.~P.~Wan, J.~X.~Wang and H.~F.~Zhang,
  ``Complete next-to-leading-order study on the yield and polarization of $\Upsilon(1S,2S,3S)$ at the Tevatron and LHC,''
  Phys.\ Rev.\ Lett.\  {\bf 112} (2014) no.3,  032001
  doi:10.1103/PhysRevLett.112.032001.

\bibitem{Han:2014kxa}
  H.~Han, Y.~Q.~Ma, C.~Meng, H.~S.~Shao, Y.~J.~Zhang and K.~T.~Chao,
  ``$\Upsilon(nS)$ and $\chi_b(nP)$ production at hadron colliders in nonrelativistic QCD,''
  Phys.\ Rev.\ D {\bf 94} (2016) no.1,  014028
  doi:10.1103/PhysRevD.94.014028.

\bibitem{Feng:2015wka}
  Y.~Feng, B.~Gong, L.~P.~Wan and J.~X.~Wang,
  ``An updated study of $\Upsilon$ production and polarization at the Tevatron and LHC,''
  Chin.\ Phys.\ C {\bf 39} (2015) no.12,  123102
  doi:10.1088/1674-1137/39/12/123102.
  
\end{thebibliography}
\end{document}